# AGILE GOVERNANCE IN INFORMATION AND COMMUNICATION TECHNOLOGIES: SHIFTING PARADIGMS

*GOVERNANÇA ÁGIL EM TIC: ROMPENDO PARADIGMAS*

**Alexandre J. H. de Oliveira Luna**
**Cleyverson P. Costa**
**Hermano P. de Moura**
**Magdala A. Novaes**
**César A. D. C. do Nascimento**
Universidade Federal de Pernambuco

___________________________________________________________________________________

## ABSTRACT

This paper presents the basis of the Agile Governance in Information and Communication Technology (ICT), which is based on Agile Software Engineering Methodologies principles and values. Its development was done through a systematic review process, supported by Bibliometrics and Scientometrics methods and techniques, where the Critical Success Factors (CSF) of ICT Governance projects and the principles of the Agile Manifesto were analyzed. Next, through an inductive approach, focused on the convergence between the concepts involved, it was analyzed how agile principles could help to minimize the gap between ICT and business. Evidences of their occurrence were taken through a Conceptual Survey Research. As a result, the foundations and concepts of

___________________________________________________________________________________

*Alexandre José Henrique de Oliveira Luna*,  PhD student and MSc in Computing Science CIn-UFPE, Center of Informatics, Federal University of  Pernambuco Post office box 7851 – 50.732-970 – Recife – PE – Brazil, Telehealth Nucleus, Department of Clinical Medicine, Group of Information Technology in Health, Clinics Hospital, Federal University of  Pernambuco, Brazil, E-mail: alexluna@mangve.org, ajhol@cin.ufpe.br

*Cleyverson Pereira Costa*, MSc in Computing Science CIn-UFPE, Center of Informatics, Federal University of  Pernambuco Post office box 7851 – 50.732-970 – Recife – PE – Brazil, E-mail: cpc@cin.ufpe.br

*Hermano Perrelli de Moura*, MSc in Informatics at Universidade Federal de Pernambuco (1989) and PhD In Computing Science at University of Glasgow (1993). Professor and Vice-Director of the Center of Informatics, Federal University of  Pernambuco Post office box 7851 – 50.732-970 – Recife – PE – Brazil, E-mail: hermano@cin.ufpe.br

*Magdala de Araújo Novaes,* PhD in Bioinformatics at Université D´Aix-Marseille II (França), Centre National de la Recherche Scientifique. Professor.  Telehealth Nucleus, Department of Clinical Medicine, Group of Information Technology in Health, Clinics Hospital, Federal University of  Pernambuco, Brazil., E-mail: magdala.novaes@nutes.ufpe.br

*César A. D. C. do Nascimento*,  Center of Informatics, Federal University of  Pernambuco Post office box 7851 – 50.732-970 – Recife – PE – Brazil, E-mail: cadcn@cin.ufpe.br

Agile Governance in ICT were defined and, finally, the development of a reference model was proposed as a future work.

**RESUMO**

Este artigo apresenta as bases do conceito de Governança Ágil em TIC – Tecnologia da Informação e Comunicação - , baseado nos princípios e valores das Metodologias Ágeis da Engenharia de Software. O desenvolvimento deste trabalho se deu através de um processo de Revisão Sistemática, apoiado em técnicas e métodos Bibliométricos e Cienciométricos, no qual foram analisados os Fatores Críticos de Sucesso (FCS) de projetos de Governança em TIC e os princípios das Metodologias Ágeis. Em seguida, através de uma abordagem indutiva, com foco na convergência entre os conceitos envolvidos, analisou-se como os princípios ágeis poderiam contribuir para minimizar o hiato existente entre a TIC e o negócio. Evidências da coerência da proposta foram reforçadas através de uma Pesquisa de Sondagem Conceitual. Como resultado, foram definidas as bases em que se fundamentam o conceito de Governança Ágil em TIC e sugerido como trabalho futuro a definição de um modelo de referência para este conceito.

**Palavras-Chave:** Tecnologia da Informação e Comunicação (TIC); Governança em TIC; Metodologias Ágeis; Gerenciamento de Projetos; Gerenciamento de Serviços de TIC.

## 1. INTRODUCTION

Currently, the corporate environment, regardless of the nature of the business and management mode (public or private), organizations are starting to realize the growing importance of Information and Communication Technology - ICT as a driving and catalyst factor of changes, renewal and achieving aspects of its business aims. In this context, institutions have increased their awareness of how ICT and its consequences have become strategic factors in increasing their market competitiveness for the achievement of its institutional mission (Eurocom, 2006).

However, for the consolidation of this strategic plan, a structured process is necessary to manage and control ICT initiatives in organizations in order to ensure the return on investments and improvements in organizational processes. In this context, the term Governance in ICT is used as a means of gaining control and knowledge in ICT, ensuring greater transparency in strategic management (Koshino, 2004).

In this way, many proposed methodologies, reference guides, sets of good practice and frameworks have emerged and thrived in recent years. This has allowed the development of ICT Governance in organizations, the rationalization of investment in ICT and metrics for assessing results in which we can highlight ITGI (2008): COBIT (ISACA, 2007) and ITIL (ITIL, 2007). However, the adoption of ICT Governance using these models, which are here called "conventional" or "traditional", does not happen without problems. Generally, these are slow processes that require high investments, which are also limited by the difficulty that organizations have understanding how to start its implementation (Magalhães and Pinheiro, 2007; Mendel and Parker, 2005; Fry, 2004; Farinha, 2005; Pegg and Kayes, 2005).

On the other hand, Agile Methodologies have spread and added increasingly competitive and dynamic approach to software development processes in Software Engineering. In this environment, independently of the business area (Luna *et al.*, 2008), we see that more and more methods for specification and software development are being gradually replaced or upgraded by the principles and values announced in the Agile Manifesto (Beck *et al.*, 2001). This occurs with the aim of obtaining results faster, and these agile methods can "add value" to business organizations, through a process in which the principles of communication and collaboration are essential (Fruhling *et al.*, 2008).

Under that view, the positive experience of organizations working in Engineering Software business was examined through a process of **Systematic Reviews** (Sampaio and Mancini, 2007) supported by **Bibliometrics** and **Scientometrics** methods and techniques (Vanti, 2002; Glänzel, 2003) . These organizations have discovered evidence of progressive and significant contributions that the Agile methods for software development processes have made (Ferreira and Lima, 2006; Dobbs, 2007; Ambler, 2007; Luna *et al.*, 2008). Based on this analysis, this paper argues for the assumption that the agile principles, values and good practices, once adapted to the context of Governance in ICT, can bring even more significant results in organizational management. Their benefits can be perceived through the increase of the speed of decision making, the insurance of business processes and the increase of organizational competitiveness and other aspects.

Thus, this proposal for Agile Governance in ICT has emerged, which provides the implementation of the principles and values of the Agile Methods to the traditional processes in ICT Governance. In a previous analysis the possibility of maximizing the potential of the critical success factors of ICT in governance through the application of the principles and values of agile methods was identified (FERNÁNDEZ, 2008), but a positive relationship in the use of an agile approach with the projects in ICT Governance is believed to be possible. This second proposal will be developed in future works.

### 1.1. Objectives

This article discusses, at a conceptual level, the critical points of conventional ICT Governance and how the principles and values of Agile Software Engineering can assist in minimizing or eliminating these problems through the identification of possible convergences between the concepts involved.

Next, this work goes beyond the theoretical approach and conducts a **Conceptual Survey** (Richardson, 1999; Marconi and Lakatos, 2004) which explores the relationship between the concepts discussed through the interview of fifty managers, professionals and post-degree students in ICT.

Finally, this paper shows the basis that underlies the concept of Agile Governance in ICT, providing the principles and values of Agile Software Engineering for the conventional Process of ICT Governance, leaving, however, the details of its application for a future work.

### 1.2. Justification

Many authors have said that in order to survive the voracity of the market, business agility is required, but what does it mean? (Scott, 2000; Roosmalen and Hoppenbrouwers, 2008; Cummins, 2008; Sloane et al., 2008). According to the Gartner Group, "business agility" is the ability to respond quickly and efficiently to changes in the business world, and transform these changes into competitive advantage (Scott, 2000).

In this context, it is observed that more organizations are adopting the agile approach as a survival tactic in these economically turbulent times (Cummins, 2008) which in turn led to interesting views. Thus, business agility is important, and according to LUFTMAN *et al.* (1993), it is the ability to "*change the direction of the environment and respond efficiently and effectively to that change*".

In essence, adding agility to the processes of Governance in ICT, already implies a higher level of convergence between ICT initiatives and business objectives which is a premise of ICT Governance. However, other benefits of an agile approach in the context of business can be identified, for example: improved time-to-market and increased speed of decision making, which ultimately reflects in increased organizational competitiveness (Roosmalen and Hoppenbrouwers, 2008).

However, establishing and extracting the best of Governance in ICT do not occur without high investments, long delays and application of models that are often beyond the needs of the organizations. Often the difficulties of putting into practice the concepts, procedures and objectives of traditional Governance Models make organizations feel powerless over the weak progress and poor visibility of the investments made.

Through an analysis of critical success factors of ICT in governance projects, and considering the principles and values from Agile Methods, it is believed that a proposed Agile Governance of ICT can help to avoid or minimize the initial errors of conventional ICT Governance, and therefore minimize the gap between ICT and business.

## 2. THEORETICAL FRAMEWORK

This article is based on two areas of influence: **Governance of ICT** and **Agile Methods**.

### 2.1. ICT Governance

**Corporate governance** is the set of processes, customs, policies, laws and institutions which affects the way a corporation is directed, administered or controlled. Corporate governance also includes the relationships between the various parties involved and the purposes for which a society is governed. The key players are the shareholders of management and board of directors. Other participants include customers, creditors (e.g. banks, holders / owners of policies / bonds), suppliers, regulators, and the wider community (Calame and Talmant, 2001).

On the other hand, **Governance of Information Technology**, **IT Governance** or **Governance in ICT**, is defined by some authors (ITGI, 2008; ISACA, 2007; ITSMF, 2008) as a subset of the corporate governance discipline, focusing on Information Technology (IT) and its performance systems and risk management. The growing interest in IT governance is partly due to the need to ensure reliable security and auditing mechanisms for companies, in order to mitigate business risk and avoid the occurrence of frauds (or ensure that there are means to identify them), ensuring transparency in management. The Sarbanes-Oxley Law (Rezzy, 2007), in the U.S., and the Basel II Accord, in Europe, are examples of mechanisms in this context. Movements such as these demonstrate how institutions that are reference in the world market recognize that ICT projects can easily get out of control and profoundly affect the performance of an organization.

With the adoption of an **ICT Governance Model,** it is expected that the structures and processes will ensure that ICT supports and maximizes the goals and strategies of the organization, allowing it to control the measurement, auditing, implementation and quality of services, and also enabling the monitoring of internal and external contracts, defining the conditions for the effective performance management based on consolidated concepts of quality. Weill and Ross (2005) state that the performance of governance evaluates the effectiveness of IT governance in meeting the four goals ranked according to their importance to the organization: i) the use of IT on a adequate cost / benefit ratio; ii) the effective use of IT for asset utilization; iii) the effective use of IT for growth; and, iv) the effective use of IT for business flexibility.

Finally, ICT Governance can be defined as the strategic alignment of ICT with the business in order to obtain the maximum value by developing and maintaining effective controls of ICT, aiming at cost control, management of return on investments and associated management risks (Weill and Ross, 2005).

To ensure such benefits, many mechanisms of relationship between business processes and ICT processes have been developed by the ICT Governance discipline. The end result of this is a plethora of standards and best practices involving: processes, indicators, profiles, guidelines, etc., whose implementation usually requires much time, money and effort, because of the formalism adopted by these standards.

Holm *et al.* (2006) present a summary of the intentions of improving the relationship between ICT and business through the classification of 17 standards and best practices in terms of the type of process and company.

This paper does not have the intention to discuss in detail the achievements or improvements that these methods and tools have achieved in order that the processes support core business of the organizations, however there is an intention to explore some context of their maximization of potentials through the new **Agile Governance in ICT** approach, as a catalyst to overcome the gap between ICT and business.

### 2.2.Agile Methodologies

**Software Engineering** (SE) has emerged as a development of Computer Science at the end of the 60s (1968), presenting itself as a proposal for the reorganization and professionalization of software development processes. This happened because of the way software projects were developed: in a disorderly, not systematic and "nearly-romantic" manner. In this way, Software Engineering is defined as an area of knowledge focused on specification, development and maintenance of software by applying technologies and practices of computer science, project management and other disciplines, aiming at organization, productivity and quality (Pressman, 2005).

Since its inception, this area is booming with the advent of several methods, techniques and tools to improve software development processes worldwide. However, even with all these developments, Software Engineering has long been facing problems related to late delivery of projects, extrapolated budgets, unsatisfied customers and users, in addition to conflict and distress among analysts and customers. This was happening, among other factors, especially because the available methods for software development were heavy, bureaucratic, inefficient and unproductive (Oliveira, 2003).

In this context, on February 11$^{th}$, 2001, a group of IT professionals and researchers met in order to start a movement towards a series of values and practices of software development which they entitled the *Manifesto for Agile Software Development* (Beck *et al.,* 2001).

They started from the premise that although each organization involved had their own practices and theories on how to make a software project succeed, each one with their special features, they all agreed that, in their previous experiences, a small set of principles always seemed to have been respected when the projects succeed. Thus, the seventeen professionals who were present signed the following:

> "*We are uncovering better ways of developing software by doing it ourselves and helping others to do so. Through this work, we have come to value:*
>
> • *Individuals and interactions over processes and tools*
>
> • *Working software over comprehensive documentation*
>
> • *Customer collaboration over contract negotiation*
>
> • *Responding to change over following a plan.*
>
> *That is, while there is value in the items on the right, we value the items on the left more.*" (Beck *et al.*, 2001).

The manifesto also sets out twelve principles of an agile process, which can be seen in Table 1.

**Table 1 – Agile Principles (BECK *et al.*, 2001).**

| ID | Principle |
|---|---|
| P1 | The **priority is customer satisfaction** through **rapid** and **continuous delivery** of software that adds **value to the business**. |
| P2 | **Changes are welcome**, even late in development, especially if the changes will provide a competitive advantage to our customers. |
| P3 | Make **frequent deliveries** of software that works from a couple of weeks to a couple of months, always looking for the shortest time between deliveries. |
| P4 | Business people (executives) and developers must **work together** daily and throughout the project. |
| P5 | Build project around **motivated individuals**. Provide all necessary support to the project environment and rely fully on the team. |
| P6 | **Face-to-face dialogue** is the most efficient and effective way to communicate the information within the development team. |
| P7 | **Software that works** is the principal measure of progress. |
| P8 | Agile processes promote **sustainable development**. The promoters, developers and users should be able *to maintain a steady work pace indefinitely*. |
| P9 | The **continuous attention** to *technical quality* and *good design* enhances agility. |
| P10 | **Simplicity is essential.** We need to know how to maximize *work that should NOT be done*. |
| P11 | The best architectures, requirements and designs emerge from **the team** itself through its **proactive** and **self-organization** (collective and collaborative intelligence[1]). |
| P12 | At regular intervals, **the team should reflect about how to become more efficient** and adjust their behavior to achieve this goal. |

In this context and seeking the best results, IT companies are adopting methodologies for developing software that are more flexible and prone to frequent changes, and more interaction throughout the project between users and the system itself. These methods are called agile methodologies as opposed to heavy methodologies that traditionally prevailed in the area, but which are inefficient and unproductive (Ferreira and Lima, 2006).

There are many agile methodologies according to Abrahamsson *et al.* (2002), all specific for project development and software maintenance. Among the most widespread agile methods we can mention XP (BECK and FOWLER, 2000) and SCRUM (Schwab and Beedle, 2002), but we can also mention: XPM - Extremme Project Management (Jacobsen, 2001), APM - agile project management (APM, 2003), fdd - feature driven development (Palmer and Felsinger, 2002), crystal family (Cockburn, 2000), DSDM - Dynamic System Development Method (Highsmith, 2002) and ASD - Adaptive Software Development (Stapleton, 1997 ), among others.

---

[1] Author´s note.

In a more comprehensive approach, allowed by the transition from these concepts of the software engineering paradigm to the governance of ICT paradigm, we can comfortably say that all these methods are based on the same vision: businesses change and users need to adapt ICT resources to these changes. This idea is crystallized in the Agile Manifesto (Beck *et al.,* 2001).

### 2.3. Limitations of ICT Governance

Once the concepts that underpin the purpose of this paper are introduced, we will present why a direct application of best practices in ICT governance is not always appropriate.

Currently, the management of ICT departments of our organizations has evolved, mostly from an ICT management system based on "firefighting" (fireman) to a state of maturity aiming at service management, as can be seen in Figure 1.

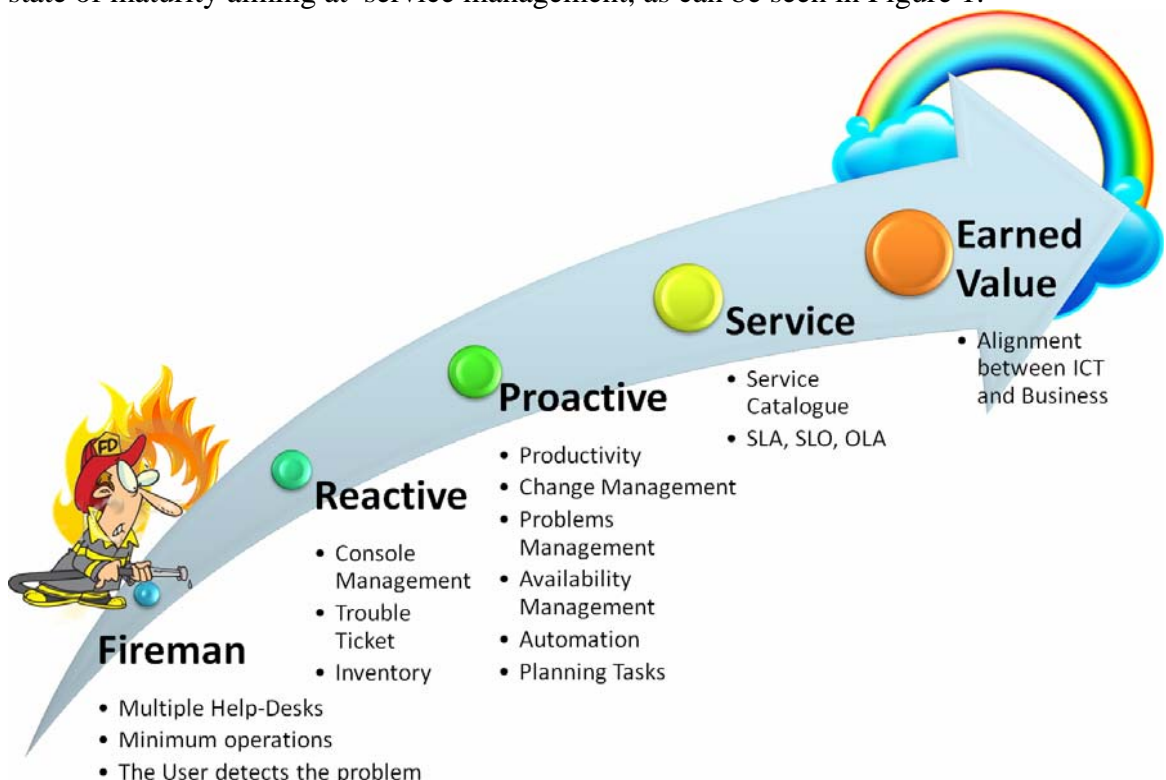

**Figure 1 – Evolution of Management of ICT departments[2].**
**Source: Adapted from (Fernández *et al*., 2008).**

But some myths about Governance in ICT need to be adressed in order to avoid the risk of failure in their adoption, such as: i) All you need to do is read all the books about ICT Governance; ii) ICT Governance tells you where to start; iii) As the ICT Governance is only a handful of books, it should be cheap; and, iv) Change Management is only for developers (Spafford and Kim, 2004).

---

[2] **Help-Desk**, the term refers to the support service to users and resolution of technical problems in computing, telephony and information technology (ITSMF, 2008).
**Trouble Ticket**, the term refers to the system registry and tracking problems in ICT in the context of an organization (ITSMF, 2008).
**SLA**, Service Level Agreement (ITGI, 2008).
**SLO**, Service Level Objective (ITGI, 2008).
**OLA**, Operational Level Agreements (ITGI, 2008).

We must not forget the fact that organizations that chose to use ICT in governance are not immune to most of the problems faced by their managers in the conduction of projects related to the subject, once they base their implementation practices on the body of knowledge Management Projects that is available, among which we highlight the PMBOK (PMI, 2004) and PRINCE2 (Bradley, 2002).

In this case, in order for ICT Governance to be effective, it must constantly analyze the level of added value to business processes, so that the governance process does not stop within itself. Thus, some important issues which organizations are facing and that can cause inefficiency in governance must be addressed strategically, such as: i) top management not seeing value in ICT investments; ii) ICT becoming a barrier to new strategic implementations for the company; iii) the mechanisms for making decisions being slow and contradictory; and iv) senior management seeing outsourcing as a solution to ICT problems (Pereira and Becerra, 2007).

One common mistake is the fact that the ICT department eventually becomes a kind of "*enlightened despotism* of the use of ICTs by the ICT" (ICT as an end in itself!), and not as a "means" to support the business needs. Below, we will comment on some erroneous approaches that are applied to some projects in ICT Governance in organizations.

1. **Too much emphasis on ICT:** One of the most common mistakes that are committed when deploying tools of ICT governance is precisely analyzing them uniquely from the technological point of view.

2. **Inherent need "to structure"**: ICT Departments are used to structuring components that are part of the ICT environment. The problem arises when this arrangement ends up forcing the creation of responsibilities, profiles, rigid hierarchical structures, overly formalized process definitions, all of which depend on agreements of static level of service. At this point, the excessive formalism can transform the entire structure into a rigid and useless model.

3. **Approaches based on general models:** there are many models to design methods and tools in ICT Governance (ITGI, 2008; Isaca, 2007, ITIL, 2007; Pereira and Becerra, 2007), but most are not specific about their implementation approach, using rather vague guidelines of "how" to apply them. Thus, generating great anxiety in the ICT Team in order to try to find out where to begin. Another aspect to be considered relates the "adherence" of the chosen model to the reality of the organization. Taking the COBIT model (ISACA, 2007), for example, which has 34 control objectives, organized and distributed in 41 documents of international character; do these 34 apply in all cases? In all organizations?

4. **Not considering people:** in an organization there are people who actually perform, control and decide about the processes, and there are business people with their decisions that aim to create value in the company. However, all the current methods and tools of ICT Governance still focus on structures and processes. There must be effective mechanisms to promote the relationship, communication and collaboration between people and organization in the context of structures and processes.

5. **The leadership of the CIO[3]:** traditionally, the figure of the CIO has presented itself as the "paladin of the causes of the ICT department", trying to defend their investments in ICT infrastructure, and acting in a tactical level, at maximum. It is necessary, however, that this character strategically repositions itself in the organization, reporting directly to the CEO[4] and supporting them in the process of strategic decision. In order for this to happen, however, it is necessary that ICT no longer be considered a center of high costs in the organization and starts to act in the strategic layer of the business, as a sector of innovation and competitive advantage.

## 3. METHODOLOGY

The definition of methodological tools is directly related to the problem being studied. The methodological framework of reference, when carefully selected, is what gives the scientific rigor of a research (Marconi and Lakatos, 2004).

According to the central goal of this research, we can classify it as **exploratory** (Gil, 2002), since it aims to present a proposal for *Agile Governance in ICT*. The preparation of the proposal was based on an **inductive** approach, supported by the procedure methods of **comparative** and **structuralist** analysis (Marconi and Lakatos, 2004). The use of such procedure methods was essential for conducting a **qualitative analysis** of the information obtained in a **bibliographic review**.

According to Gil (2002), **exploratory studies** are intended to provide greater familiarity with the subject, aiming to make it more explicit or to form hypotheses. You could say that these researches aim at the improvement of ideas or the discovery of intuitions. Most often this type of research takes the form of **literature review** or **case study** (Gil, 2002).

The method of **inductive approach** is characterized by using a set of private data, enough to infer a general truth, not necessarily contained in the parts examined. Its application is divided into three steps: i) observation of phenomena; ii) discovery of the relationship between them; and, finally iii) the generalization of the findings (Marconi and Lakatos, 2004).

The review process can occur through **systematic reviews**, as well as through other types of review studies. Systematic review is a form of research that uses as the literature from a particular theme a source. This type of research provides a summary of evidence related to a specific intervention strategy, by applying systematic and explicit methods of researching, critical appraisal and summary of selected information (Sampaio and Mancini, 2007).

---

[3] *Chief Information Officer*
[4] *Chief Executive Officer*

The potential of this systematic review method may be maximized by the application of **Bibliometrics** and **Scientometrics** techniques and methods, in order to ensure relevance of the selected material in the literature for the review process (Vanti, 2002; Glänzel, 2003).

For the exploration of the relationship between the concepts discussed, the **Conceptual Survey Research** (Richardson, 1999; Marconi and Lakatos, 2004) was applied to a group of fifty managers, professionals and graduate ICT students. The questionnaires were processed; the outcome was examined and compared to the results of a preliminary analysis.

In this work, this combination of methods and techniques has been applied in several crucial moments, which can include: review of the state of the art of Governance in ICT, review of agile methodologies, review of scientific work in theoretical and practical implementation and improvement of Governance in ICT, among others.

For this research, a comprehensive review was initially conducted on the state of the art of the concepts involved, through a process of Systematic Reviews supported by the Bibliometrics and Scientometrics techniques and methods. In a second stage, the identification of critical success factors of ICT Governance projects and analysis of its adherence to the principles of the Manifesto of Agile Software Engineering was started.

Then, the **Conceptual Survey Research** (Richardson, 1999) was applied, in which a questionnaire, based on two tables, each one containing 12 items, was presented. In the first table, the 12 Agile Principles, called simply "Table A - Principles" were presented. In the second Table, called "Table B – Factors", The Critical Success Factors of Project Governance were presented. However an explanation over what the two tables were about, or what their meaning was, in order to avoid biased answers in completing the questionnaire, was not given. The participants were asked to fill a 12x12 matrix, resulting from the combination of all the principles of Table A with the factors in Table B, seeking to relate the concepts involved in every possible combination, according to **Table 2**.

**Table 2 – Criterion score of the responses.**
**SOURCE: Adapted from Likert scale (Richardson, 1999).**

| Punctuation | Signification |
|:---:|---|
| 3 | Very convergent |
| 2 | Convergent |
| 1 | Slightly convergent |
| 0 | No relationship |
| -1 | Slightly divergent |
| -2 | Divergent |
| -3 | Very divergent |

The principles of the punctuation used were based on the associative logic of Likert Scale (Richardson, 1999). This scale is a type of psychometric response scale, commonly used in questionnaires, and in most opinion polls. When responding to a questionnaire based on this scale, the level of agreement with a given statement is evaluated. In this research, the scale has seven points of agreement, according to **Table 2.** An additional order was also given, when relationships of divergence were identified, in addition to registering the negative score, the respondents should also write a justification for it.

The forms received were tabulated on a spreadsheet that was pre-formatted according to the following considerations:

i. Three levels of consideration for the responses were established, in light of the identification and characterization information, which follows: (weight 1) Graduate Students in ICT (Weight 2), Professionals working in the ICT area, and (Weight 3) Managers working with ICT.
ii. After classifying the person, according to the previous item, the response was arithmetically weighed according to the equivalent profile, generating a Weighted Array of Single Response (WASR).
iii. The WASR was then weighed, originating the Unified Response Matrix (URM).
iv. Some points of difference identified were analyzed in light of the justification presented. If any inconsistency was identified in the justification, the entire questionnaire was discarded.

## 4. RESULTS

Before discussing the results of the Conceptual Survey Research, we shall consider the preliminary analysis derived from systematic reviews based on scientific literature about the concepts involved, and what motivated the early stages of exploration.

In this way, considering the critical points presented in the application of conventional ICT Governance, an approach to pontentialize the critical success factors of ICT Governance was tried, seeking to answer the following question: Which is the appropriate approach to reduce the failure of the deployments and operations of ICT Governance projects?

Aiming to answer this question, the research was initiated by analyzing the aspects that could influence the successful adoption of ICT Governance in organizations. According to Albertin (2004), **Critical Success Factors – CSF-** are factors that, if not considered and managed, inevitably undermine the success of the initiative.

Thus, after examining the most common **Critical Success Factors** in the literature (Mezzomo and Pasqualetti, 2006; Albertin, 2004), and crossing this information with actual case studies about adoption of ICT in governance, according to some authors (Pereira, 2007; Techrepublic, 2002; Techrepublic, 2003; Holm *et al.,* 2006), **Table 3 was obtained,** with the critical success factors identified in the most relevant projects on ICT Governance, categorized according to Albertin (2004).

**Table 3 – Critical Success Factors of ICT Governance Projects.**

| ID | Factor | Categorization according to (Albertin, 2004) |
|---|---|---|
| F1 | From a Reference Model (a framework), you can deploy enterprise processes in an organization or even in a sector (according to their needs). | • Organization<br>• Planning<br>• Control |
| F2 | Maintenance of operational procedures based on the "deliveriables" of each part of this framework. | • Control |
| F3 | Acculturation of all employees of the corporation, before the day-to-day changes that occur with the implementation of operational processes. | • People<br>• Organization |
| F4 | Involvement of the senior management of the organization is essential to the sponsorship of the decisions and priorities of projects. | • Organization<br>• Sponsoring |
| F5 | Involvement of all stakeholders and those affected by the practices of governance implemented in the organization. | • People |
| F6 | Existence of process change management and internal dissemination (*endomarketing*[5]) to minimize internal resistance. | • Organization<br>• People |
| F7 | Focus on small, consecutive victories and disseminating the results of initiatives. | • Control<br>• People |
| F8 | Frequent and constant communication of progress during implementation. | • Organization<br>• People<br>• Planning |
| F9 | Planning and managing the project scope. | • Planning |
| F10 | Caution in the deployment process of simultaneous innovations, minimizing the risk of not meeting the initial objectives outlined. | • Organization<br>• Planning<br>• Control |
| F11 | Using the existing organizational infrastructure to accelerate the process. | • Organization<br>• Sponsoring |
| F12 | Considering the process of continuous improvement of ICT Services. | • Control<br>• Sponsoring |

The factors cited refer to the organization, to the users and the methodology. Thus, we can conclude, firstly, that an approach to *Agile Governance in ICT* focused on user needs should be promoted.

In this sense, could this new proposal be based on Agile Methods? The answer could be affirmative if there was a relationship between the critical factors of success and the agile principles outlined above in **Table 1**.

In this analysis, a matrix of relationships was developed, demonstrating the positive relations with a "+", the negative with a "-" and the null with "blank space". We obtained a profile of the relationship between these two aspects, as can be seen in **Table 4**, according to preliminary analysis performed by comparing the convergence of the concepts covered.

---

[5] It tries to adapt strategies and elements of traditional marketing, normally used in the external organizations, for use in an internal corporate environment (RICHARDSON, 1999).

The reasoning applied on the identification of convergence or divergence between the agile principles in **Table 1** and the critical success factors in **Table 3**, took into consideration the perception of "**cohesion**" or "**distance**" of the concepts involved in the generation of **Table 4**. Aiming to illustrate the rational process applied, the ordered pair P6 x F8 in **Table 4** was selected for exemplification:

  i.    **P6:** The "face-to-face dialogue" is the most efficient and effective way to communicate the information within the development team.
  ii.   **F8:** "Frequent and constant communication" of progress during implementation.
  iii.  **Reasoning**: Both coordinates approach the aspect of "communication" as essential to their paradigms, showing a clear convergence relationship.
  iv.   **Result:** "+", positive.

**Table 4 – Relationship between Critical Success Factors of Projects of ICT Governance and the Agile Principles.**

| Principle / Factor | P1 | P2 | P3 | P4 | P5 | P6 | P7 | P8 | P9 | P10 | P11 | P12 |
|---|---|---|---|---|---|---|---|---|---|---|---|---|
| F1  | + | + |   | + | + |   |   | + | + | + | + | + |
| F2  |   |   | + | + | + |   | + | + | + |   | + | + |
| F3  | + | + | + | + | + | + |   | + |   | + |   | + |
| F4  | + |   |   | + | + |   |   |   |   |   | + |   |
| F5  |   |   |   | + | + | + |   | + |   |   | + | + |
| F6  | + | + |   | + | + |   | + | + |   |   | + |   |
| F7  | + | + | + | + | + |   |   | + | + | + | + | + |
| F8  | + |   |   | + | + | + | + |   |   |   | + | + |
| F9  |   | + | + |   |   |   | + | + | + | + | + |   |
| F10 | + |   |   |   |   |   |   | + | + | + | + | + |
| F11 | + | + |   | + | + |   |   | + |   |   | + | + |
| F12 | + |   | + | + |   |   |   | + | + | + | + | + |

From **Table 4** we can deduce that there is an apparent positive relationship in using an agile methodological approach in addressing governance projects in ICT. Naturally, it is possible to extrapolate these results and point out an approach to Agile Governance in ICT based on the values and principles of the Agile Manifesto, and with the intention to avoid the initial errors already existing in conventional ICT Governance.

Based on the principles of the inductive process used for the relationship between Agile Principles and Critical Success Factors of Governance Project, a **Conceptual Survey Research** was conducted (Richardson, 1999; Marconi and Lakatos, 2004), aiming to confirm and consolidate the "inferred positive relationship" between these two areas of knowledge.

After processing the survey questionnaires, the following Unified Response Matrix (URM) was obtained, as can be seen in **Table 5**. The range of colors represented in **Table 5** shows the numeric ranges used in **Table 2**.

**Table 5– Unified Response Matrix. SOURCE: Own elaboration.**

| Unified Response Matrix (URM) | | | | | | | | | | | | |
|---|---|---|---|---|---|---|---|---|---|---|---|---|
| Principle / Factor | P1 | P2 | P3 | P4 | P5 | P6 | P7 | P8 | P9 | P10 | P11 | P12 |
| F1 | 1,24 | 1,01 | 1,36 | 1,35 | 0,97 | 0,78 | 1,74 | 1,59 | 1,40 | 1,04 | 1,06 | 1,27 |
| F2 | 1,08 | 0,73 | 1,35 | 0,90 | 0,86 | 0,72 | 1,59 | 1,46 | 1,62 | 0,90 | 1,05 | 1,09 |
| F3 | 1,10 | 1,72 | 1,26 | 1,62 | 1,56 | 1,40 | 1,27 | 1,49 | 1,62 | 1,18 | 1,65 | 1,82 |
| F4 | 1,82 | 1,77 | 1,08 | 1,90 | 1,54 | 1,65 | 1,37 | 1,56 | 1,45 | 1,21 | 1,32 | 1,64 |
| F5 | 1,68 | 1,85 | 1,26 | 1,85 | 1,83 | 1,68 | 1,49 | 1,62 | 1,73 | 1,41 | 1,82 | 1,91 |
| F6 | 1,12 | 1,97 | 0,95 | 1,21 | 1,22 | 1,14 | 1,17 | 1,27 | 1,10 | 1,03 | 1,22 | 1,36 |
| F7 | 1,31 | 1,06 | 1,50 | 1,23 | 1,72 | 1,18 | 1,32 | 1,24 | 1,46 | 0,91 | 1,36 | 1,44 |
| F8 | 1,72 | 1,76 | 1,68 | 1,76 | 1,73 | 1,88 | 1,40 | 1,54 | 1,41 | 1,36 | 1,56 | 1,77 |
| F9 | 1,74 | 1,71 | 1,59 | 1,65 | 1,36 | 1,10 | 1,55 | 1,46 | 1,29 | 1,47 | 1,38 | 1,31 |
| F10 | 1,42 | 1,06 | 1,36 | 1,06 | 1,08 | 0,67 | 1,50 | 1,42 | 1,51 | 1,24 | 1,12 | 1,26 |
| F11 | 1,03 | 0,88 | 1,37 | 1,26 | 1,35 | 0,78 | 1,41 | 1,53 | 1,26 | 0,72 | 1,32 | 1,32 |
| F12 | 1,68 | 1,41 | 1,36 | 1,62 | 1,41 | 1,23 | 1,81 | 1,71 | 1,73 | 1,08 | 1,31 | 1,56 |

Applying a 3D surface graph to the MRU data in **Table 5**, **Figure 2 was obtained**. This chart illustrates the degree of convergence between the concepts involved based on the survey.

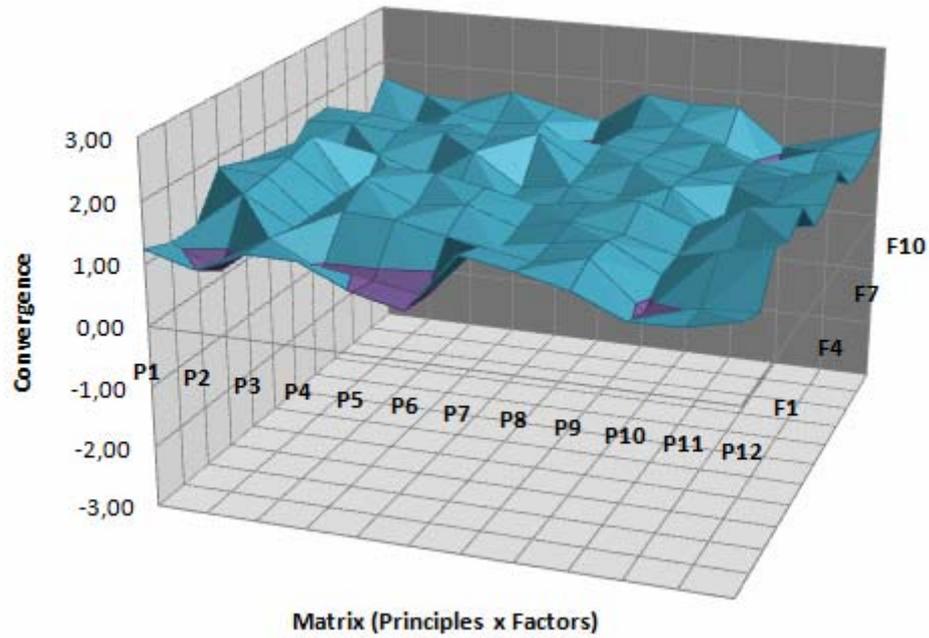

Figure 2.(A) Y - axis angle of 20 degrees

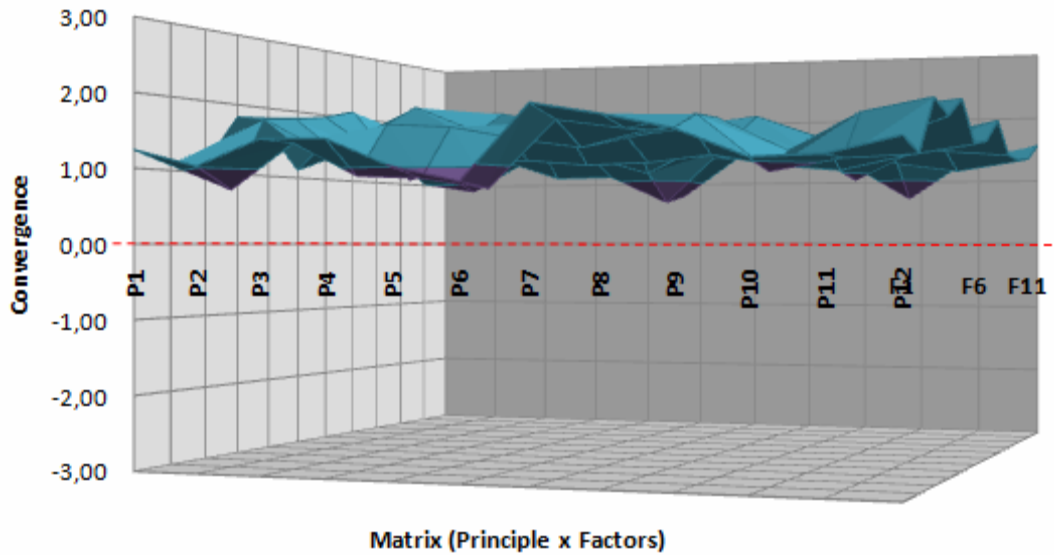

Figure 2.(B) Y - axis angle of zero degrees

**Figure 2– Curve surface of the relationship between the concepts of the tables in Research. SOURCE: Own elaboration.**

Analyzing the results of the Conceptual Survey Research, it was observed that the "apparent" positive relationship deduced subjectively at the beginning of this section, and inferred the completion of **Table 4**, is confirmed statistically by **Table 5** and **Figure 2**, once all points on the surface curve in **Figure 2-B** are in the positive quadrant of 3D chart.

Based on the results of the research conducted, a position of greater security was assumed, in order to suggest the adoption of an agile methodological approach to support ICT Governance projects.

It is also possible to speculate that this approach can be taken without the need to develop a new method of Governance in ICT. It might be enough just to adjust the focus of existing ones, such as COBIT (ISACA, 2007) and ITIL (ITIL, 2007) to agree with the principles and values supported by the Agile Manifesto, and with the application of good practices that can be adapted from the Agile Software Engineering.

Usually almost all the initiatives in ICT Governance occur through the implementation of projects. Regardless of the nature of the organization and business, the project-based culture is increasingly rooted in the corporate environment, according to the results that have been obtained with the application of methodologies, reference guides, best practices and professionalism in this area of knowledge (PMI, 2004).

When the organization identifies that it is time to go toward Governance in ICT, this is executed through projects of implementing the processes of governance. Insofar as governance processes are gradually implemented and come into operation, it is necessary to manage ICT services, which have to be monitored as operations through its SLAs (Service Level Agreements) agreed with the various parties involved. Still, when changes or improvements are required in these cases or wish to offer other ICT services to the organization, often these changes occur also through projects (Weill e Ross, 2005), (ITSMF, 2008).

In this case, projects end up being the vehicle through which initiatives of ICT Governance are lead, as well as the management of changes that arise from them.

Thus, it is believed that once the essence of the principles, values and practices of the Agile Paradigm from Software Engineering are translated into the context of Governance in ICT, the basis of Agile Governance in ICT will be prepared. It is also believed that it is possible to develop this model as a practical guide for implementing agile principles and values, using projects as the "vehicle" for this change / transformation, and getting, as a result, the desired reduction in the gap between ICT and business in organizations.

Using the definitions of ICT Governance and having found that the most appropriate methodological approach points to an agile orientation, the first definition of this new concept is based on the use of joint ownership and adding value to business, concepts covered by the principles of the Agile Manifesto, in which all Agile Methods are based. Thus, the concept of Agile Governance in ICT is proposed, such as:

> "**Agile Governance in ICT** *is the process of defining and implementing the ICT infrastructure that will provide support to strategic business objectives of the organization, which is jointly owned by ICT and the various business units and instructed to direct all involved in obtaining competitive differential strategic <u>through the values and principles of the Agile Manifesto</u>"*.

To facilitate the understanding of the concept of *Agile Governance in ICT,* a brief comparison of some relevant aspects in the implementation of the different approaches used has been drawn up in **Table 6**. This table compares the aspects of focus, language and relationships of the different approaches: traditional ICT Management, ICT Governance, Agile Methods and Agile Governance in ICT. The approach of Agile Governance in ICT can be observed as a result of the convergence of aspects of ICT Governance and Agile Methods in contrast to conventional ICT Management.

**Table 6 – Comparative analysis of different approaches about the focus, language and relationship. SOURCE: Own elaboration.**

| ID | Aspects | Traditional ICT Management | ICT Governance | Agile Methods | *Agile Governance in ICT* |
|---|---|---|---|---|---|
| 1 | Focus | On the Technology | On the Business | On the Customer | On the business of the Customer |
| 2 | Language | Technological | Business | Customer | Business Customer |
| 3 | Relationship with the Customer | Limited and distant | Participative | Close | Close and participative, acting with the Customer to decide the factors that enables agility to the process. |

In this way, **Figure 3** aims to present the inter-relationship between these areas of knowledge. This figure suggests that through the extraction of principles, values and best practices of the Agile Paradigm of Software Engineering, and focusing on aspects related to deployment and process improvement, added to the context of Governance in ICT which is embedded in traditional ICT Management, it is possible to construct the ICT *Agile Governance* concept.

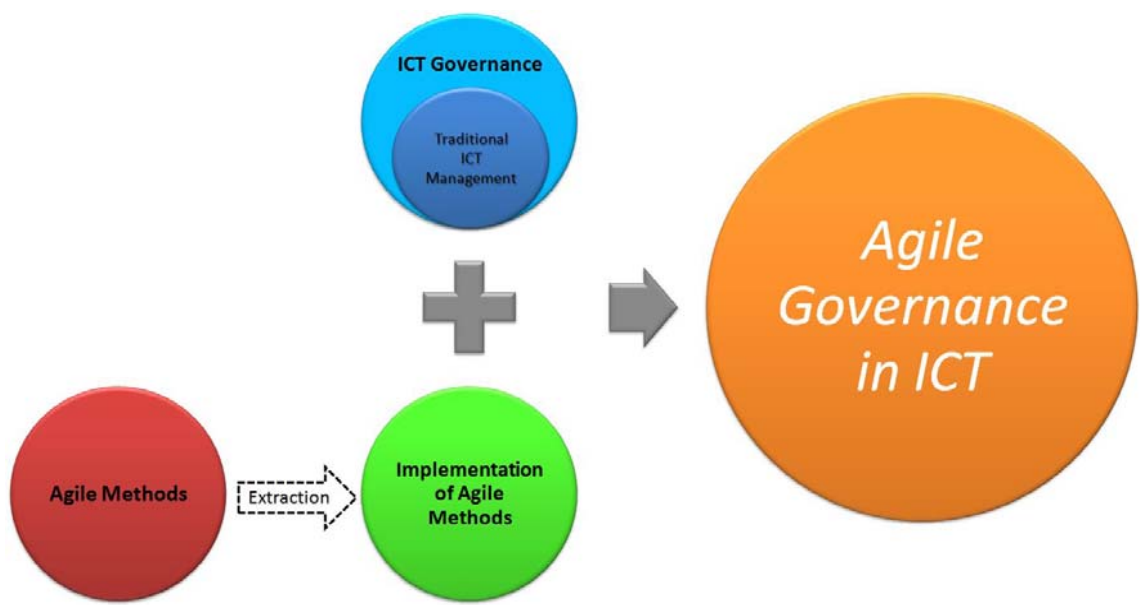

**Figure 3 – Diagram of the interrelationship between the knowledge areas involved. SOURCE: Own elaboration.**

## 5. Conclusions and Perspectives

This article presented the basis on which the concept of **Agile Governance in ICT** rests, as a result of operating the positive relationship identified between the concepts of the principles of *Agile Manifesto of Software Engineering* (BECK *et al.*, 1999) and the *Critical Success Factors of Projects* of implementation and improvement of Governance in ICT (MEZZOMO and PASQUALETTI, 2006; ALBERTIN, 2004).

In this context, this paper proposes as its main **contribution**, the conduction of a **conceptual survey**, used to corroborate the preliminary results obtained by analyzing the information extracted from systematic reviews about the concepts involved. Also, it presents the **definition of the term Agile Governance in ICT** and the **exploitation of the convergence between the concepts addressed**. Both the **motivation to move forward** and develop a tool or a method for agile approach to Governance in ICT, as well as the submission of a proposal by the **positioning of this new concept** in relation to areas of knowledge that it deals with, as detailed by **Table 6** and **Figure 3,** can also be considered as a contribution of this work.

As **constraints**, the fact that the Conceptual Survey Research has been applied to a small, though qualified universe, can be cited. Another restriction of the work could be the difficulty finding consistent scientific publications and studies about the implementation and improvement of processes and services of ICT Governance. Likewise, the case studies found, followed little scientific rigor most of the time, which reduces the potential of the results.

As **future work**, we propose the creation of a reference model for this concept, the identification of the paradigms to shift, the components of governance, the decision-making areas, the roles, relationships, actions that promote the elimination or minimization the gap between ICT and business. This reference model must be validated in real organizations, and should be flexible enough to have adhesion (coupling) to the most well known Governance in ICT models in the market.

**Acknowledgments**

Special thanks to *Carmen Raquel Nunes Silva* for revising the translation of the article into English.